\documentstyle[12pt,epsf,psfig]{article}
\begin{document}
\baselineskip 0.6cm
\newcommand{\gsim}{ \mathop{}_{\textstyle \sim}^{\textstyle >} }
\newcommand{\lsim}{ \mathop{}_{\textstyle \sim}^{\textstyle <} }
\newcommand{\vev}[1]{ \left\langle {#1} \right\rangle }
\newcommand{\bra}[1]{ \langle {#1} | }
\newcommand{\ket}[1]{ | {#1} \rangle }
\newcommand{\EV}{ {\rm eV} }
\newcommand{\KEV}{ {\rm keV} }
\newcommand{\MEV}{ {\rm MeV} }
\newcommand{\GEV}{ {\rm GeV} }
\newcommand{\TEV}{ {\rm TeV} }
\newcommand{\nn}{\nonumber}


\begin{titlepage}

\begin{flushright}
UT-923\\
\end{flushright}

\vskip 2cm
\begin{center}

{\Large \bf Constraints on 
Inflation Models \\ from Supersymmetry Breakings}

\vskip 1.2cm
Koshiro Suzuki and T.~Watari

\vskip 0.4cm
{\it Department of Physics, University of Tokyo, \\
         Tokyo 113-0033, Japan}\\

(\today)

\vskip 1.5cm
\abstract{ Effects of soft SUSY-breaking terms on inflation potentials
are discussed. There exist generic constraints that must be satisfied in
order not for the inflaton potential to loose its flatness. We examine
explicitly the constraints in the case of a hybrid inflation model and
find that the coupling constant $\lambda$ between the inflaton and the
``water-fall-direction'' field is bounded as $2.0 \times 10^{-6} \lsim
\lambda$. This is a highly non-trivial result. Indeed, if we adopt the
severest constraint from avoiding the problem of the gravitinos produced
non-thermally, $\lambda \lsim 7.4 \times 10^{-6}$ is required, under a
reasonable assumption on the reheating process. This means that the
hybrid inflation model marginally has a viable parameter space. We also
discuss analogous constraints on other inflation models.}

\end{center}
\end{titlepage}


Supersymmetry (SUSY) is not only motivated from the low-energy particle
physics, but also from the primordial inflation. SUSY naturally explains
the existence of the inflaton scalar field, and moreover, the
cancellation of radiative corrections in supersymmetric theories is
preferable for maintaining the flatness of the inflaton potential. There
have been constructed various models for inflation in the framework of
supergravity such as chaotic \cite{chaotic}, hybrid \cite{hybrid}, new
\cite{new}, topological \cite{topological}, D-term \cite{D-term}
inflation models and so on.
 
Each model, in general, has some free parameters even after the
COBE-normalization condition 
\begin{equation}
\frac{\delta \rho}{\rho} = \frac{1}{5\sqrt{3}\pi}
\frac{ V^{3/2}(\varphi_{ N_{e}} ) }{M_P^3 | V'(\varphi_{N_{e}}) |} 
  \simeq \frac{1}{5\sqrt{3}\pi} \times 5.3 \times 10^{-4}
\label{eq:COBE-normalization}
\end{equation}
is imposed. Here, $M_P$ denotes the reduced Planck scale $2.4 \times
10^{18}\GEV$, $V$ the inflaton potential, $V^{'}$ the derivative of the
potential with respect to the inflaton $\varphi$, and $\varphi_{N_e}$
the value of the inflaton field at the time the COBE-scale exited the
horizon.  Different values of model parameters lead to different
reheating processes after those inflations.  Successful inflation models
must have viable parameter spaces whose resulting reheating processes
satisfy all phenomenological constraints such as those from the
gravitino problem.

Whether a sufficient e-fold number is obtained for a given model
parameter must be examined using the {\it full} supergravity potential;
it must be noted that the inflation sector is not the only one which
couples to supergravity. In particular, the hidden sector, which has
SUSY-breaking expectation values, is also coupled to supergravity. Then,
the inflaton potential receives soft SUSY-breaking contributions, as in
the ordinary gravity mediation of the SUSY breaking to the
standard-model sector\footnote{Ref.\cite{soft} pointed out that these
SUSY breaking terms can have physical importance.}. It is true that the
gravity-mediated soft SUSY-breaking terms appear with
$1/M_P$-suppression and are tiny. However, the flatness of the
slow-roll-inflaton potential requires a tuning at the level of order of
the Hubble parameter $H = \sqrt{V/3}/M_P$, which is also a
Planck-suppressed quantity. Therefore, the perturbations to the inflaton
potential by soft SUSY-breaking terms can be crucial for its
flatness. This observation gives new constraints on the viable parameter
space of successful inflation.

In this letter, we show how the existence of the soft SUSY-breaking
terms constrains the parameter spaces of inflation models. We apply the
above-mentioned new generic constraints to a hybrid inflation model in
supergravity, as an example, and derive a lower bound on the Yukawa
coupling of the inflaton field $\varphi$ and the inflaton mass
$m_\varphi$. We show that this new lower bound is highly significant if
we take seriously the severest constraint from avoiding the gravitino
problems. We also discuss analogous constraints on other inflation
models.
 
The slow-roll inflation works well when the slow-roll conditions 
of the inflaton potential,
\begin{eqnarray}
 \epsilon &\equiv& \left( \frac{V'(\varphi)M_{P} }{V(\varphi)} \right)^{2} \ll 1,
\label{eq:epsilon} \\
 \eta &\equiv& \frac{V''(\varphi) M_{P}^{2} }{V(\varphi)}, \hspace{1cm} |\eta| \ll 1, 
\label{eq:eta} 
\end{eqnarray}
are satisfied \cite{Kolb:1990vq}. Gravity-mediated soft SUSY-breaking
terms should not violate the above two conditions. We can robustly say
from the condition eq.(\ref{eq:eta}) that
\begin{equation}
 |m_{\rm S\not{U}SY}^2| \lsim |V^{''}| \ll \frac{V}{M_P^2} \simeq H^2,
\label{eq:generic-constraint}
\end{equation}
where $m_{\rm S\not{U}SY}$ is the soft SUSY-breaking mass.  This
condition must be satisfied for any inflation that obtains not-so-small
e-fold number,\footnote{It is possible that the soft SUSY-breaking terms
accidentally cancel the supersymmetric contribution to the inflaton
potential and do not ruin the flatness.  In this special case, this
generic constraint is not necessarily applicable.} and puts a lower
bound on the possible energy scale of inflations for a fixed gravitino
mass. We can obtain severer constraints for each model, although they
are model dependent. In the following, we take the hybrid inflation
model as an example. Constraints on other inflation models are discussed
later.

The hybrid inflation models in supergravity are given by the
following K\"ahler potential and superpotential \cite{hybrid}:
\begin{eqnarray}
K &=& \Phi^\dagger \Phi - \frac{k}{4}\frac{(\Phi^\dagger \Phi)^2}{M_P^2} 
  + \cdots 
\label{kahler_hybrid} \\
W &=& \Phi (\lambda \Psi \bar{\Psi} - \Lambda_I^2 ).
\label{super_hybrid}
\end{eqnarray} 
Here, the radial part of the scalar component of the $\Phi$ multiplet
($\varphi$) plays the roll of the inflaton. $\Lambda_I$ is the energy
scale of the inflation ($V \simeq \Lambda_I^4$). The SUSY contribution
to the inflaton potential is given by 
\begin{equation}
 V(\varphi) = \Lambda_I^4 \left( 1 
                     + \frac{\lambda^2}{8\pi^2}\ln (\frac{\varphi}{\varphi_c}) 
                     + \frac{k}{2}\left(\frac{\varphi}{M_P}\right)^2 
                     + \cdots \right),
\label{eq:hybrid-potential}
\end{equation} 
where $\varphi_c \equiv \sqrt{2/\lambda}\Lambda_I$ is the value of the
inflaton field where the slow-roll inflation ends. The second term comes
from the 1-loop renormalization.

Now we expect that the soft SUSY-breaking terms exist,
\begin{equation}
V_{\rm S\not{U}SY}(\varphi) \simeq \frac{1}{2}m_{\rm S\not{U}SY}^2\varphi^2 
 + |A| \Lambda_I^2 \varphi ,
\end{equation}
where $m_{\rm S\not{U}SY}$ is the soft SUSY-breaking mass and $A$ is a
dimension-one SUSY-breaking parameter. Combined with the SUSY
contribution in eq.(\ref{eq:hybrid-potential}), effective mass squared
of the inflaton is given by $ (\Lambda_I^4/M_P^2) k + m_{\rm
S\not{U}SY}^2$.
One thing we notice is that we can change the quadratic term in the
total inflation potential by changing the value of $k$ only if
\begin{equation}
\left| \left(\frac{m_{\rm S\not{U}SY}^2}{\Lambda_I^4/M_P^2} \right) \right| \lsim k.
\label{eq:lower-bound-k}
\end{equation}
Another thing we point out here is that the slow-roll condition of
$\eta$ is always violated unless
\begin{equation}
 \left| m_{\rm S\not{U}SY}^2 \right| M_P^2  \ll \Lambda_I^4,
\label{eq:m-SUSY-Li}
\end{equation}
which is nothing but eq.(\ref{eq:generic-constraint}).
The other slow-roll condition of $\epsilon$ is also violated unless
\begin{equation}
 |A|^2 M_P^2  \ll \Lambda_I^4.
\label{eq:A-Li}
\end{equation}
If the soft SUSY-breaking mass and the A-parameter are given by the
gravitino mass as in the gravity-mediation model, the energy scale of
the inflation is bounded from below as
\begin{equation}
(m_{3/2} M_P)^2 \ll \Lambda_I^4.
\label{eq:lower-bound-Li}
\end{equation}
Again, this bound must be satisfied by any inflation which occurred in
the thermal history, irrespective of whether it satisfies the
COBE-normalization condition eq.(\ref{eq:COBE-normalization}) or not.
That is, this bound is also
applicable to inflations that occurred later than the COBE-normalized
inflation \cite{later-inflation}. The energy scale of any inflation must
be higher than the SUSY-breaking scale $\Lambda_{S\not{U}SY} \sim
\sqrt{m_{3/2}M_P}$.

A few comments on the induced soft SUSY-breaking terms are in order here.

We assume that the soft SUSY-breaking contributions to the inflation
potential during the inflation is not so different from those expected
from the present vacuum of the hidden sector. Suppose the hidden sector
has flat directions. Then it is possible that the field expectation
values of the hidden sector fields are not settled at the present value,
since eq.(\ref{eq:generic-constraint}) holds. In those cases, there is a
possibility that the condensation energy of the hidden sector is by far
different from the present value, leading to soft SUSY-breaking terms
drastically different from those induced by the present vacuum of the
hidden sector. However, this possibility is out of the scope of the
discussion in the following.

The other limitation of our analysis is that we assume $m_{\rm
S\not{U}SY} \sim |A| \sim m_{3/2}$ (gravity-mediation model). For the
light gravitino mass $m_{3/2} \lsim 10 \GEV$ (gauge-mediation model),
the inflaton can acquire soft SUSY-breaking terms whose soft parameters
are of order of the weak scale $( \gg m_{3/2})$, if the inflatons are
charged under the standard-model gauge group. For heavy gravitino mass
$10 \TEV \lsim m_{3/2}$ (anomaly-mediation model), the inflaton obtains
only loop-suppressed soft SUSY-breaking terms, if the inflaton sector
is, along with the observed sector, suitably separated from the hidden
sector in the K\"ahler potential. We also neglect these cases, although
necessary modification for these two models of SUSY-breaking mediation
is straightforward.

Now, let us see explicitly how the generic constraints discussed above
put limits on the parameters of the hybrid inflation model. We apply the
constraint
eq.(\ref{eq:lower-bound-k})\footnote{Eq.(\ref{eq:lower-bound-Li}) is
satisfied if we use $k \lsim 1$ and eq.(\ref{eq:lower-bound-k}).}  to
the inflation that satisfies the COBE-normalization condition
eq.(\ref{eq:COBE-normalization}). The field value $ \varphi_{N_e}$ in
eq.(\ref{eq:COBE-normalization}) is determined by requiring the obtained
e-fold number to be suitable:
\begin{equation}
67 + \frac{1}{3} \ln \left(\frac{H T_R}{M_P^2} \right) = N_{e} = \frac{1}{M_P^2}\int_{\varphi_{c}}^{\varphi_{N_e}} 
             d \varphi \; \frac{V(\varphi)}{V'(\varphi)} ,
\label{eq:e-fold-67}
\end{equation}
where $H$ is the Hubble parameter during the inflation and $T_R$ the
reheating temperature after the inflation. The hybrid inflation model
given by eqs.(\ref{kahler_hybrid}) and (\ref{super_hybrid}) has
three parameters, namely $\Lambda_I, k$ and $\lambda$, among which
$\Lambda_I$ can be expressed as a function of $k$ and $\lambda$ due to
the COBE-normalization condition eq.(\ref{eq:COBE-normalization}).

\begin{figure}[ht]
\centerline{\psfig{figure=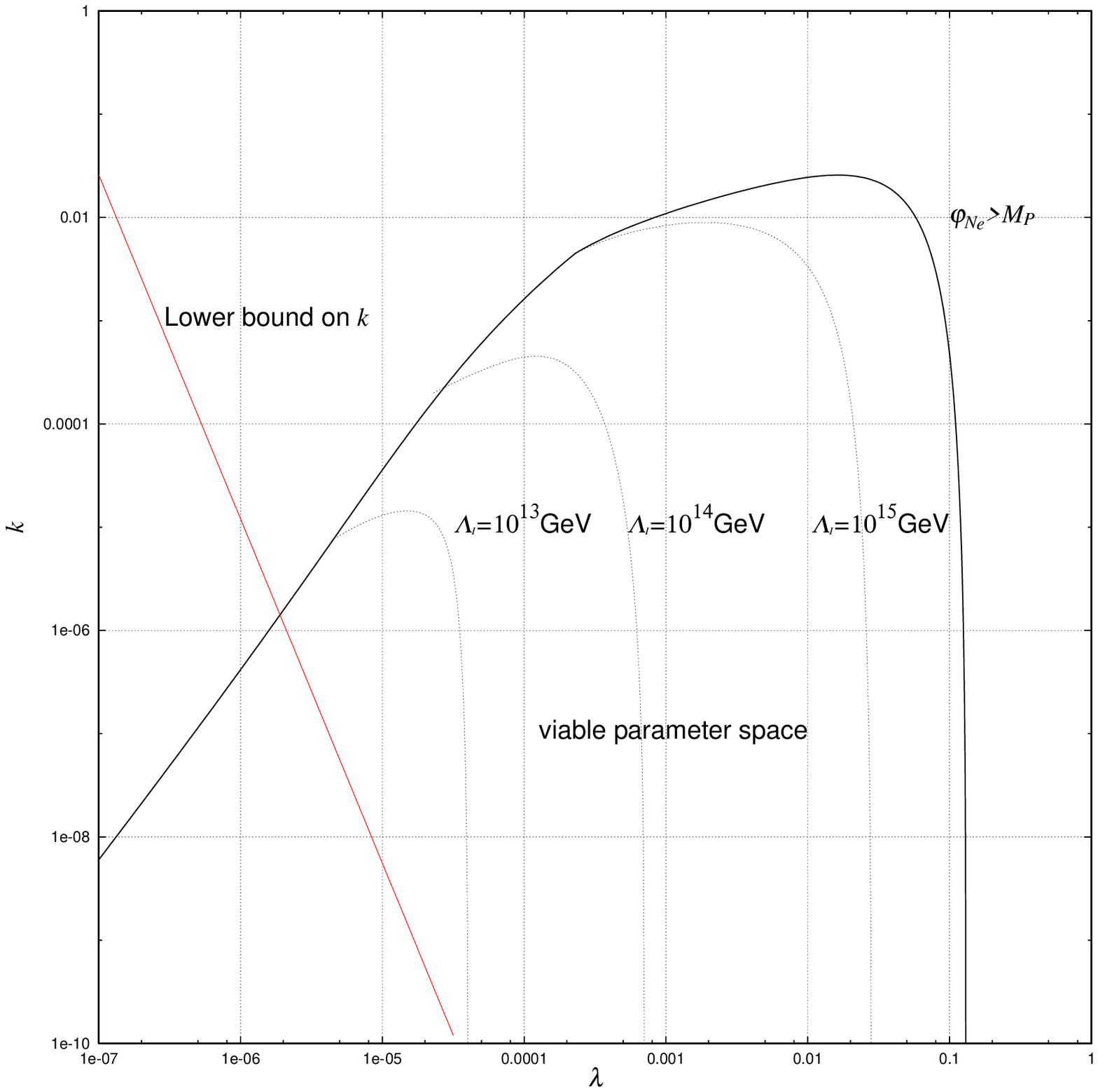,width=15cm}} 
\caption{The viable parameter space of the COBE-normalized hybrid
inflation model. The red (dashed) line is the lower bound on $k$ from
the SUSY-breaking soft terms. The thick black solid line is the upper
bound on $k$ from the requirement $\varphi_{N_{e}} < M_{P}$. We can see
from this that $\lambda$ is lower bounded, $\lambda \gsim 2.0 \times
10^{-6}$. We also show the contours for fixed $\Lambda_I$. The lines
except the red (dashed) line are cited (with some modification) from
Ref.\cite{Lepto}. We thank the authors for permission.}
\label{fig:hybrid}
\end{figure}

It was shown in Ref.\cite{Lepto} that the suitable e-fold number is
obtained for the parameter space below the thick black solid line
described in Fig.\ref{fig:hybrid},\footnote{Assumption on the reheating
process does not change this result very much.}which was derived in the
absence of the soft SUSY-breaking terms. Furthermore, they showed
numerically that for smaller $\lambda$ and smaller $k$, the energy scale
of the inflation $\Lambda_I$ becomes lower.

We present an analytical argument and visualize what happens when the
Yukawa coupling $\lambda$ becomes smaller.  For sufficiently small
$\lambda \ll 3 \times 10^{-3}$ with negligible $k$,\footnote{This
parameter region corresponds to the situation where the following two
conditions are satisfied: $(1)~$ $\varphi_c \sim \varphi_{N_e} \ll M_P$
holds, and $(2)~$ $ V'(\varphi) \simeq
\Lambda_I^4\left((\lambda^2/8\pi^2)(1/\varphi) + (k\varphi
/M_P^2)\right)$ is dominated by the previous term.}or precisely
speaking,
\begin{equation}
 2.2 \times 10^{1} \lambda^{\frac{4}{3}} = 
 2.2 \times 10^{-7} \left(\frac{\lambda}{10^{-6}}\right)^{\frac{4}{3}}
 \gg k,
\label{eq:negligible-k}
\end{equation}
$\Lambda_I$ and $\varphi_{Ne}$ are solved in terms of $\lambda$ and $k$
through eq.(\ref{eq:COBE-normalization}) and eq.(\ref{eq:e-fold-67}) as:
\begin{equation}
  \left( \frac{\Lambda_I}{M_P}\right) \simeq 1.7 \times 10^{-2} 
       \lambda^{\frac{5}{6}},\label{eq:Li-dependence}
\end{equation}
and\footnote{In eq.(\ref{eq:pn-pf-dependence}), we calculate the
reheating temperature using eq.(\ref{eq:decay-rate}) as the decay rate
of the inflaton. This assumption does not make any essential difference
in the following analysis. See also footnote 4.}
\begin{eqnarray}
 \left(\frac{\varphi_c}{M_P}\right)^2 &\simeq& 5.8 \times 10^{-4} 
       \lambda^{\frac{2}{3}} , \label{eq:pf-dependence}\\
 \left(\frac{\varphi_{N_e}-\varphi_c}{M_P}\right) &\simeq& 0.53 
    (61 + \frac{3}{4} \times \ln \lambda) \lambda^{\frac{5}{3}}.
                               \label{eq:pn-pf-dependence}
\end{eqnarray}
Notice that the energy scale of the inflation is determined only by
$\lambda$ because we consider negligibly small $k$ [see
eq.(\ref{eq:negligible-k})]. The exponents of $\lambda$ in
eqs.(\ref{eq:Li-dependence})--(\ref{eq:pn-pf-dependence}) can be easily
understood by observing that the following three relations must be
satisfied:
\begin{equation}
\left(\frac{\varphi_c}{M_P} \propto \lambda^{\frac{1}{3}}\right)
  = \sqrt{\frac{2}{\lambda}}\left(\frac{\Lambda_I}{M_P}\right)
  \propto \lambda^{-\frac{1}{2}} \lambda^{\frac{5}{6}}
  \propto \lambda^{\frac{1}{3}},
\end{equation}
\begin{equation}
 \left( N \sim 60 + {\rm factor} \times \ln \lambda \right)
 \propto \frac{1}{\lambda^2} \left(\frac{\varphi_{N_e}-\varphi_c}{M_P}\right)
                              \left(\frac{\varphi_c}{M_P}\right)
 \propto \lambda^{-2} \lambda^{\frac{5}{3}} \lambda^{\frac{1}{3}} 
 \propto \lambda^0,
\end{equation}
\begin{equation}
\left( \frac{\delta \rho}{\rho} \sim 10^{-5}\right)
   \propto \left(\frac{\Lambda_I}{M_P}\right)^2 \frac{1}{\lambda^2}
           \left(\frac{\varphi_c}{M_P}\right)
   \propto \lambda^{\frac{5}{3}} \lambda^{-2} \lambda^{\frac{1}{3}} 
   \propto \lambda^0.
\end{equation}
After all, we see that $\Lambda_I$ decreases as $\Lambda_I \propto
\lambda^{5/6}$ when $\lambda$ decreases.

Remember that $\Lambda_I$ is bounded from below
[see eq.(\ref{eq:lower-bound-Li})] because of the existence of the soft
SUSY-breaking mass terms. This means that there exists a lower bound on
the coupling $\lambda$. Indeed, eq.(\ref{eq:lower-bound-k}), which says
that the parameter $k$ is bounded from below, is now expressed as (for
$\lambda \ll 3 \times 10^{-3}$ and eq.(\ref{eq:negligible-k}))
\begin{equation}
 k \gsim 
   \left( \frac{M_P}{\Lambda_I}\right)^4 \left(\frac{m_{3/2}}{M_P}\right)^2 
   \simeq 
   1.2 \times 10^{-5} \left( \frac{m_{3/2}}{240 \GEV}\right)^2
                      \left( \frac{\lambda}{10^{-6}}\right)^{-\frac{10}{3}}.
\label{eq:lower-bound-k2} 
\end{equation}
This constraint eq.(\ref{eq:lower-bound-k2}) is described in
Fig.\ref{fig:hybrid} as the red (dashed) line.  The region between the
red (dashed) line and the thick black solid line in Fig.\ref{fig:hybrid}
is the viable parameter space of the hybrid inflation model with a
suitable e-fold number, after we take the SUSY-breaking terms into
account.  It can be seen from the figure, as we stated before, that the
possible value of $\lambda$ has a lower bound:
\begin{equation}
  \lambda_{\rm min.} \simeq 2.0 \times 10^{-6} 
                    \left(\frac{m_{3/2}}{240 \GEV}\right)^{\frac{3}{7}} 
     \lsim \lambda, 
\label{eq:lower-bound-l}
\end{equation} 
where we have used an empirical relation read off from the figure
\begin{equation}
 4.9 \times 10^{-7}\left( \frac{\lambda}{10^{-6}}\right)^{\frac{4}{3}} \gsim k
 \hspace{1cm} ({\rm for} \;\; 10^{-6} \lsim \lambda \lsim 10^{-4}).
\end{equation}

The energy scale of the inflation at the extreme case of $\lambda \simeq 
\lambda_{\rm min.}$ in eq.(\ref{eq:lower-bound-l}) is
\begin{equation}
\Lambda_I \simeq 
  3.0 \times 10^{-7} \left(\frac{m_{3/2}}{240 \GEV}\right)^{\frac{5}{14}}M_P 
= 7.2 \times 10^{11} \left(\frac{m_{3/2}}{240 \GEV}\right)^{\frac{5}{14}} \GEV,
\end{equation}
which is slightly larger than the energy scale of the 
SUSY breaking, $\Lambda_{\rm S\not{U}SY} \equiv
3^{1/4} \sqrt{m_{3/2}M_P} = 3.2 \times
10^{10} \times (m_{3/2}/240 \GEV)^{1/2}\GEV$, and hence we see 
that the soft SUSY-breaking terms become important for the smallest
value of $\lambda$ obtained in eq.(\ref{eq:lower-bound-l}). 

Now let us briefly discuss the implication of the lower bound on the
parameter $\lambda$. The lower bound eq.(\ref{eq:lower-bound-l}) is far
below the {\it natural} value (order one), or far below the upper bound
for a sufficient e-fold number $\lambda \lsim (1-2) \times 10^{-1}$
\cite{Lepto}. Therefore, the bound seems to have almost no physical
importance at first sight. However, a sufficient e-fold number is not
the only requirement to the inflation model; reheating after the
inflation must lead to a suitable initial condition of the Big-Bang
cosmology, and this requirement gives much stronger constraint on
$\lambda$.

For instance, the requirement from the problem of thermally-produced
gravitinos puts an upper bound on the parameter $\lambda$ \cite{Lepto},
under an assumption that the decay rate of the inflaton is given by
\begin{equation}
\Gamma = \frac{N}{8\pi}\frac{m_\varphi^3 
\left(\frac{\Lambda}{\sqrt{\lambda}}\right)^2}{M_P^4},
\label{eq:decay-rate}
\end{equation} 
where $N$ is the number of the decay modes of the inflaton (of order of
100), and $m_\varphi \equiv \sqrt{2\lambda} \Lambda_I$ the inflaton mass.  The inflaton mass $m_\varphi$
is proportional to $\lambda^{4/3}$, and the reheating temperature is
proportional to $\lambda^{7/3}$. It is shown in Ref.\cite{Lepto} that
$\lambda \lsim 10^{-3}$ is necessary for $T_R \lsim 10^{6}\GEV$. This
upper bound on $T_R$ is derived to avoid the thermal overproduction of
gravitinos \cite{HKKM}.

Recently, it has been discussed that the gravitinos can also be produced
non-thermally during the matter-dominated era before the reheating, and
that the abundance of these gravitinos can be larger than that of the
gravitinos produced thermally after the reheating. There are two
different estimations of the yield ({\it i.e.} the ratio of the number
density $n_{3/2}$ to the entropy $s$) of the non-thermally produced
gravitinos \cite{gravitino-non-thermal,gravitino-non-thermal-2}. Among
these two, the estimation of Ref.\cite{gravitino-non-thermal-2} predicts
a larger value of the yield, and hence leads to a stronger constraint on
the successful inflation models.
 The estimation is given by%
\footnote{While preparing this paper for publication, we received the
preprint \cite{Nilles} which says that the gravitinos produced while the
Hubble parameter is larger than the gravitino mass remain as inflatinos
under the time evolution and do not turn into the present day
gravitinos in a model they considered.
In the case this statement also holds in a hybrid inflation model, 
the yield of the non-thermally produced gravitinos reduces significantly from
eq.(\ref{eq:non-thermal}).}
\begin{equation}
\frac{n_{3/2}}{s} \sim 10^{-2} \frac{m_\varphi^3 T_R}{(m_{3/2}M_P)^2}.
\label{eq:non-thermal}
\end{equation}
Under the assumption of the inflaton decay rate
eq.(\ref{eq:decay-rate}), eq.(\ref{eq:non-thermal}) leads to
\begin{equation}
 \frac{n_{3/2}}{s} \sim 3.1 \times 10^{-12} \left(\frac{M_P}{m_{3/2}}\right)^2
\lambda^{\frac{19}{3}}
\end{equation}
for $\lambda \ll 3 \times 10^{-3}$ and negligibly small $k$ ({\it
i.e.} eq.(14)). The cosmological upper bound on the yield of the
gravitino ($n_{3/2}/s \lsim 10^{-12}$ \cite{HKKM}) leads to the
following upper bound on $\lambda$;
\begin{equation}
\lambda \lsim 7.4 \times 10^{-6} \left(\frac{m_{3/2}}{240\GEV}\right)^{\frac{6}{19}}.
\label{eq:lambda-upper-bound}
\end{equation} 
Therefore, even if we adopt the severest estimation of the non-thermal
production of the gravitinos, there exists a narrow (but not vanishing)
parameter space between the upper bound from avoiding the gravitino
overproduction in eq.(\ref{eq:lambda-upper-bound}) and the lower bound
from the soft SUSY-breaking terms eq.(\ref{eq:lower-bound-l}).  In
conclusion, the hybrid inflation model has a viable parameter space
where sufficient e-fold number is obtained and at the same time where
the yield of the gravitinos produced thermally and non-thermally is
sufficiently suppressed.

%

We have pointed out that the existence of the soft SUSY-breaking terms
from the hidden sector can violate the required flatness of the
inflation potential, and hence can restrict the viable parameter space
of the model. Our consideration has led to the existence of a lower
bound on the parameter $\lambda$ in a hybrid inflation model. Analyses
for other inflation models are straightforward, and we do not repeat
them here; we only describe the results. The consequences of the
gravitino problem can also be derived easily.

For chaotic inflation model \cite{chaotic}, our discussion does not give
any new information, because the COBE-normalization condition sets the
inflaton mass to be of order of $10^{13}\GEV$, which is far above the
soft SUSY-breaking mass.


For the new inflation model in the 2nd reference of \cite{new},%
\footnote{See Ref. \cite{new} for the notations.}%
we need to consider two cases, according to which term dominates in the
first derivative of the inflation potential $V'(\varphi_{N_e})$, {\it
i.e.}, the $\varphi^1_{N_e}$ term or the $\varphi^{n-1}_{N_e}$ term.  In
the $\varphi^1_{N_e}$-dominant parameter region ($1/[(n-2)N_e + (n-1)] <
k$), the energy scale of the inflation $\Lambda_I$ is written in
terms of $k$ and $g$ as
\begin{equation}
\Lambda_I \simeq M_P \cdot (\sqrt{2} c)^{\frac{n-2}{2(n-3)}} \cdot
 k^{\frac{n-1}{2(n-3)}} \cdot (ng)^{-\frac{1}{2(n-3)}} \cdot e^{-\frac{n-2}{2(n-3)}kN},
\end{equation}
where $c \equiv 5.3 \times 10^{-4}$ in eq.(\ref{eq:COBE-normalization}).
%
Our constraint eq.(\ref{eq:generic-constraint}) gives a lower bound on
$k \Lambda_I^4(k,g)$ for a fixed gravitino mass and fixed $k$, and as a
consequence, $g$ is bounded from above for fixed $k$ (since $\Lambda_I
\propto g^{-\frac{1}{2(n-3)}}$). Then, the inflaton mass $m_{\varphi}
\propto g^{- \frac{2}{n(n-3)}}$ is bounded from below for fixed $k$, 
and this lower bound takes its minimum value at $k \simeq 0.1$;%
%
\footnote{The observational constraint on the spectral index $n_s \gsim
0.8$ puts an upper bound on $k$ (since $n_s=1-2k$).}%
it is of order of $10^7 \GEV$ for $n=4$.
%
In the $\varphi_{N_e}^{n-1}$-dominant parameter region ($0 < k <
1/[(n-2)N_e + (n-1)]$), the inflaton mass is always larger than the
smallest inflaton mass in the $\varphi_{N_e}^1$-dominant parameter
region.

For the new inflation model in the 1st reference of \cite{new}, the
potential is the same as that discussed in the previous paragraph, with
one extra constraint between $\Lambda_I$ and $g$ imposed.  As a result,
we can write $g$ (and hence the inflaton mass $m_{\varphi} \simeq n
g^{1/n} \Lambda_I^{2(n-1)/n} / M_P^{(n-2)/n}$) in terms of $k$ and $n$
(in the $\varphi_{N_e}^1$-dominant region), or $n$ (in the
$\varphi_{N_e}^{n-1}$-dominant region). We see that the lower bound on
the inflaton mass is of order of $10^7 \GEV$, again at the observed
upper bound on $k$, although this bound has nothing to do with the
existence of the soft SUSY-breaking terms in this model.
%


For the topological inflation model \cite{topological}, the energy scale of
inflation $\Lambda_I$ can be written solely in terms of $\kappa \equiv
2g + k -1$, where $g$ is the Yukawa coupling in the superpotential and
$k$ is the four-point coupling in the K\"ahler potential; $\Lambda_I =
(c\kappa)^{1/2} M_{P} e^{- \kappa N / 2}$, where $c \equiv 5.3\times
10^{-4}$.
Our constraint eq.(\ref{eq:generic-constraint}) gives a lower bound on
$\kappa \Lambda_I^4(\kappa)$ for a fixed gravitino mass, and hence a
bound on $\kappa$, $10^{-8} \lsim \kappa \lsim 0.5$. However,
eq.(\ref{eq:generic-constraint}) does not give any constraint on $g$, so
thermal and non-thermal production of gravitinos can be sufficiently
suppressed by taking $g$ small enough.
\footnote{Inflaton mass $m_{\varphi}$ is written as $m_{\varphi} = 2
\sqrt{g}\Lambda_I^2 /M_P$ in this model.}%
%

For a D-term inflation model \cite{D-term}, the potential of this model
is nothing but that of the hybrid inflation model
eq.(\ref{eq:hybrid-potential}) with $k=0$, $\lambda$ replaced by the
gauge coupling $g$, and $\Lambda_{I}^4$ by $\frac{1}{2}(g^2 \xi^4)$,
where $\xi^2$ is the Fayet-Iliopoulos parameter. The same analysis
developed in this paper is applicable, and the gauge coupling $g$ is
bounded from below. However, the inflaton mass still depends upon the
Yukawa coupling $\lambda$ in the superpotential, which is not
constrained by the presence of the hidden sector. Therefore, no new
results on the reheating process are derived.

\section*{Acknowledgments}
We thank Tsutomu Yanagida for his careful reading of this manuscript,
precious comments, and continuous encouragements. T.W. thanks the Japan
Society for the Promotion of Science for financial support.

\newpage

%
%
%
\newcommand{\Journal}[4]{{#1} {\bf #2}, {#3} {(#4)}}
\newcommand{\PL}{\sl Phys. Lett.}
\newcommand{\PR}{\sl Phys. Rev.}
\newcommand{\PRL}{\sl Phys. Rev. Lett.}
\newcommand{\NP}{\sl Nucl. Phys.}
\newcommand{\ZP}{\sl Z. Phys.}
\newcommand{\PTP}{\sl Prog. Theor. Phys.}
\newcommand{\NC}{\sl Nuovo Cimento}
\newcommand{\MPL}{\sl Mod. Phys. Lett.}
\newcommand{\PRep}{\sl Phys. Rep.}

%

\begin{thebibliography}{99}
%
%
%
%
\bibitem{chaotic}
H.~Murayama, H.~Suzuki, T.~Yanagida and J.~Yokoyama, 
\Journal{Phys.Rev.}{D50}{2356-2360}{1994};\\
M.~Kawasaki, Masahide Yamaguchi and T.~Yanagida,
\Journal{Phys.Rev.Lett.}{85}{3572-3575}{2000}.  
%
%
\bibitem{hybrid}
	E.~J.~Copeland, A.~R.~Liddle, D.~H.~Lyth, E.~D.~Stewart and D.~Wands,
	   \Journal{Phys.Rev.}{D49}{6410}{1994};\\
        A.D.Linde, \Journal{Phys.Lett.}{B259}{38}{1991};\\
        G.~Dvali, R.~Schaefer and Q.~Shafi,
           \Journal{Phys.Rev.Lett.}{73}{1886}{1994};\\
        G.~Lazarides, R.~Schaefer and Q.~Shafi,
           \Journal{Phys.Rev.}{D56}{1324}{1997};\\
        A.~Linde and A.~Riotto, \Journal{Phys.Rev.}{D56}{1841}{1997}.
%
%
\bibitem{new}
K.-I.~Izawa and T.~Yanagida, \Journal{Phys.Lett.}{B393}{331-336}{1997};\\ 
Ref.\cite{Lepto}.
%
%
\bibitem{topological}

K.I.~Izawa, M.~Kawasaki and T.~Yanagida, 
\Journal{Prog.Theor.Phys.}{101}{1129-1133}{1999};\\ 
M.~Kawasaki, N.~Sakai, Masahide Yamaguchi and  T. Yanagida, 
\Journal{Phys.Rev.}{D62}{123507}{2000}. 
%
%
\bibitem{D-term}
E.~Stewart, \Journal{Phys.Rev.}{D51}{6847}{1995};\\
E.~Halyo, \Journal{Phys.Lett.}{B387}{43}{1996};\\
P.~Bin\'etruy and G.~Dvali, \Journal{Phys.Lett.}{B388}{241}{1996}.
%
%
\bibitem{soft}
W.~Buchm\"uller, L.~Covi and D.~Del\'epine,
 \Journal{Phys.Lett.}{B491}{183-189}{2000}. 
%
%
%
\bibitem{Kolb:1990vq}
E.~W.~Kolb and M.~S.~Turner,
``The Early Universe,''
{\it  Redwood City, USA: Addison-Wesley (1990) 547 p. (Frontiers in physics, 69)}.
%
%
\bibitem{later-inflation}
M.~Kawasaki, N.~Sugiyama and T.~Yanagida, 
\Journal{Phys.Rev.}{D57}{6050-6056}{1998}; \\
T.~Kanazawa, M.~Kawasaki, N.~Sugiyama and T.~Yanagida, 
\Journal{Phys.Rev.}{D61}{023517}{2000};\\ 
T.~Kanazawa, M.~Kawasaki, N.~Sugiyama and T.~Yanagida, 
astro-ph/0006445. 
%
%
%
%
%
%
%
\bibitem{Lepto}
T.~Asaka, K.~Hamaguchi, M.~Kawasaki and T.~Yanagida, 
\Journal{Phys.Rev.}{D61}{083512}{2000}. 
%
%
\bibitem{HKKM}
   E.~Holtmann, M.~Kawasaki, K.~Kohri, T.~Moroi,
        \Journal{Phys.Rev.}{D60}{023506}{1999}. 
%
%
\bibitem{gravitino-non-thermal}
G.~Giudice, A.~Riotto and I.~Tkachev, 
\Journal{JHEP}{9908}{009}{1999},\Journal{JHEP}{9911}{036}{1999}. 
%
%
\bibitem{gravitino-non-thermal-2}
D.~H.~Lyth,
\Journal{Phys.Lett.}{B488}{417-422}{2000}, 
\Journal{Phys.Lett.}{B476}{356-362}{2000};\\ 
H.~B.~Kim and D.~H.~Lyth, hep-ph/0011262.
%
%
%
%
\bibitem{Nilles}
   H.P.~Nilles, M.~Peloso and L.~Sorbo, hep-ph/0102264.	
%
%
\end{thebibliography}
\end{document}